\documentclass[11pt,a4paper]{article}
\pdfoutput=1
\usepackage{jheppub}

\usepackage{amsfonts}
\usepackage{slashed}

\def\bea{\begin{eqnarray}}
\def\eea{\end{eqnarray}}
\def\nn{\nonumber}

\def\lmatrix{\left(\begin{array}}
\def\rmatrix{\end{array}\right)}

\def\msbar{\overline{\rm MS\kern-0.5pt}\kern0.5pt}
\def\ig{\includegraphics}

\title{More on the flavor dependence of $m_\varrho / f_\pi$}

\author[a]{Andrey Yu. Kotov,}
\author[b]{Daniel Nogradi,}
\author[ac]{Kalman K. Szabo}
\author[b]{and Lorinc Szikszai}

\affiliation[a]{Julich Supercomputing Centre, Forschungszentrum Julich, D-52425, Germany}
\affiliation[b]{E\"otv\"os University, Institute for Theoretical Physics, Budapest 1117, Hungary}
\affiliation[c]{Department of Physics, Wuppertal University, Gaussstr. 20, D-42119, Germany}

\emailAdd{a.kotov@fz-juelich.de}
\emailAdd{nogradi@bodri.elte.hu}
\emailAdd{k.szabo@fz-juelich.de}
\emailAdd{szikszail@caesar.elte.hu}

\abstract{In previous work, arXiv:1905.01909, we have calculated
the $m_\varrho / f_\pi$ ratio in the chiral and continuum limit for $SU(3)$ 
gauge theory coupled to $N_f = 2,3,4,5,6$ fermions in the fundamental representation. 
The main result was that this ratio displays no statistically significant $N_f$-dependence.
In the present work we continue the study of the $N_f$-dependence by extending the simulations
to $N_f = 7, 8, 9, 10$. Along the way we also study in detail the $N_f$-dependence of finite volume
effects on low energy observables and a particular
translational symmetry breaking unphysical, lattice artefact phase specific to staggered fermions.}


\arxivnumber{}

\begin{document}

\maketitle

\section{Introduction and summary}
\label{intro}

We study the flavor number dependence of the ratio of the vector meson mass and the pseudoscalar decay
constant in $SU(3)$ gauge theory. The ratio is significant for a large class of beyond Standard Model
theories envisioning a strongly interacting Higgs sector and a composite Higgs boson \cite{Bardeen:1989ds}. 
The elementary fermion
ingredients of the composite Higgs boson may form other bound states, such as a vector meson, which would
be one of the new, so far undetected, particles the theory predicts. The pseudoscalar decay
constant sets the scale, in many theories it is simply identified with $v = 246.22\;$GeV, the symmetry
breaking scale of the Standard Model. Having non-perturbative results for $m_\varrho / f_\pi$ then
determines the vector meson mass $m_\varrho$ in physical units. This beyond Standard Model scenario, and
variants thereof, attracted enormous interest in the lattice community in the past decade
\cite{Fodor:2009wk,DelDebbio:2010hu,DelDebbio:2010hx,Bursa:2011ru,Fodor:2012ty,Appelquist:2013pqa,Aoki:2013zsa,Hietanen:2014xca,Appelquist:2014zsa,Aoki:2014oha,Aoki:2015jfa,DeGrand:2015lna,DelDebbio:2015byq,Aoki:2016wnc,Appelquist:2016viq,Fodor:2016wal,DeGrand:2016htl,Arthur:2016dir,Fodor:2016zil,Appelquist:2017vyy,Appelquist:2017wcg,Ayyar:2017qdf,DelDebbio:2017ini,Appelquist:2018yqe, Ayyar:2018glg,Ayyar:2018zuk,Appelquist:2019lgk,Fodor:2019vmw,Brower:2019oor,Ayyar:2019exp,Fodor:2020niv,Appelquist:2020xua,Hasenfratz:2020ess}. For a recent
reviews of the available lattice results see \cite{Drach:2020qpj} and references therein.

Apart from the phenomenological motivation the $N_f$-dependence of our ratio is an interesting QFT
question on its own. Once both $f_\pi$ and $m_\varrho$ are understood to be defined at finite fermion mass $m$
and the chiral limit is only taken for the ratio, $m_\varrho / f_\pi$ is a meaningful quantity both
inside and outside the conformal window. Outside the conformal window both the denominator and nominator
are finite in the chiral limit with an obviously finite ratio.
Inside the conformal window both $m_\varrho$ and $f_\pi$
behave as $O(m^\alpha)$ for small $m$ with the same exponent $\alpha$, again leading to a finite ratio
in the chiral limit. Hence the ratio is meaninful and well-defined on the full range $0 \leq N_f \leq 16$, 
including the quenched case $N_f = 0$ and the last integer flavor number $N_f = 16$ before asymptotic freedom is
lost at $N_f = 33/2$. Formally, $N_f = 33/2$ corresponds to a free theory \cite{Cichy:2008gk} 
and as such $m_\varrho = 2m$
and $f_\pi = \sqrt{12} m$, leading to $m_\varrho / f_\pi = 1 / \sqrt{3}$. This is an order of magnitude
smaller than $\sim 8$ found for $2 \leq N_f \leq 6$. Hence on the range $7 \leq N_f \leq 16$
the ratio will drop an order of magnitude and it is not a priori known
whether the drop will be gradual or rapid, nor is it known if the onset of the conformal window somewhere
around $10 \leq N_f \leq 13$ is connected to it in any way.

Motivated by both the phenomenological implications and the purely QFT aspects we continue the
investigation with $7 \leq N_f \leq 10$ in the present work. Even though we would like to know the
behavior for $11 \leq N_f \leq 16$ as well, finite volume effects are growing as a function of $N_f$ so
rapidly that unfortunately we must postpone these flavor numbers to future work.

The organization of the paper is as follows. In section \ref{discretization} we first study the
$N_f$-dependence of an unphysical lattice phase specific to staggered fermions. The reason for doing so
is that as $N_f$ grows the size of the unphysical phase in the $(\beta,m)$ plane grows and one must avoid
it in order to perform the physically relevant chiral-continuum limit. Section \ref{finitevolumeeffects}
details our study of the finite volume effects, the upshot of which is that as $N_f$ is growing so do
finite volume effects. In fact the growth is rather rapid and is the main reason $N_f = 10$ is the
highest flavor number we can reliably simulate at the moment. The chiral-continuum limit is
investigated in section \ref{chiralcontinuum} once the bare parameters are chosen such that unphysical phases are
avoided and finite volume effects are suppressed sufficiently. We end with conclusions and possible
outlook to future work in section \ref{conclusion}.

\section{Discretization and unphysical phases with staggered fermions}
\label{discretization}

The lattice discretization in the present work follows exactly \cite{Nogradi:2019iek}; 4 steps of stout smearing
\cite{Morningstar:2003gk, Durr:2010aw}
is applied to naive staggered fermions with smearing parameter $\varrho = 0.12$. A combination of the HMC
and RHMC algorithms \cite{Duane:1987de, Clark:2006fx} 
with or without rooting are used to have the desired continuum flavor number $N_f$.

Both in \cite{Nogradi:2019iek} and the present work simulations are run at
particular points of the $(\beta,m)$ phase diagram at given $N_f$. It is            
important that the bare parameters are all in the region of phase space which is continuously connected
to the physical $\beta\to\infty$ region, especially because unphysical phases do exist with staggered
fermions.

\begin{figure}
\begin{center}
\includegraphics[width=7.5cm]{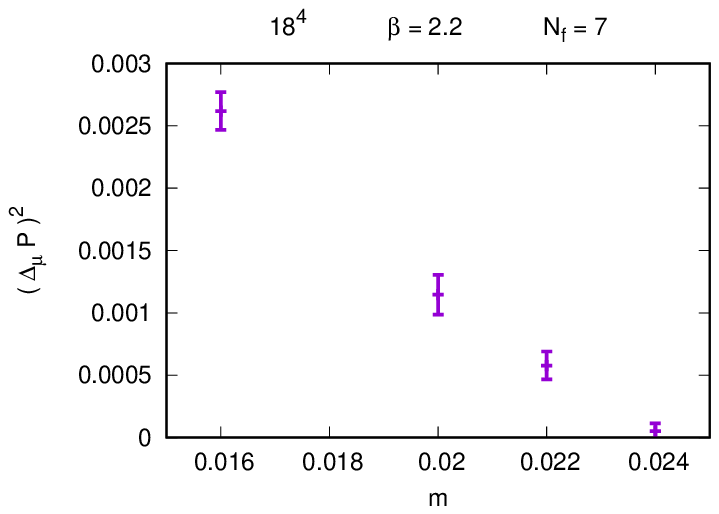} \includegraphics[width=7.5cm]{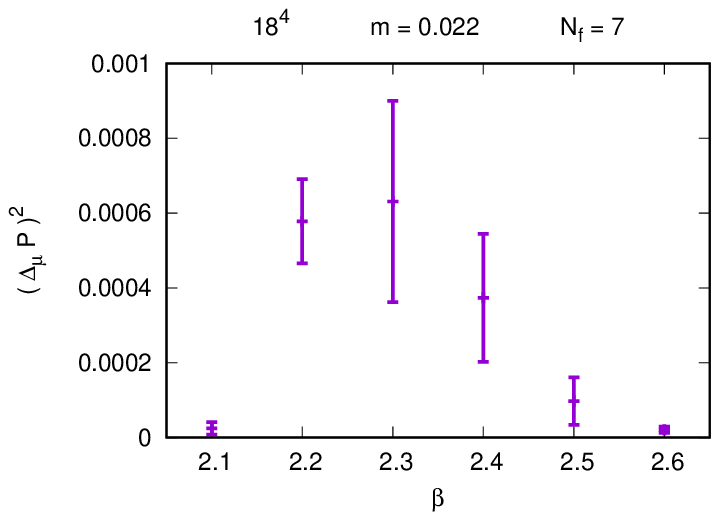} 
\end{center}
\caption{Two examples at $N_f = 7$ for finding the boundary of the shift symmetry broken phase. The
    square of the observable (\ref{defp}) is shown at fixed $\beta$ (left) and fixed $m$ (right). }
\label{deltap}
\end{figure}

The possibility that in the $(\beta,m)$ bare parameter space an unphysical Aoki-like phase might exist
with staggered fermions was first pointed out in \cite{Lee:1999zxa, Aubin:2004dm}. 
Using staggered chiral perturbation theory it was
shown that decreasing the mass on coarse lattices can lead to condensation of taste split meson states which in turn
means that the vacuum becomes unstable. The new vacuum has different symmetries from the one expected in the continuum 
and in particular the so-called staggered shift symmetry, which is a translation by a single site
accompanied by a phase factor for fermion fields, is broken.
Briefly, taste split
meson masses $M^2$ in staggered chiral perturbation theory receive a continuum-like contribution 
from the fermion mass, $O(m)$, but also a contribution from taste splitting operators, 
$O(a^2)$. If the latter is negative and large in absolute value compared to the former, 
$M^2$ may turn negative, leading to the aforementioned instability.
Convincing numerical evidence for
this scenario was provided in \cite{Cheng:2011ic} for $N_f = 8, 12$ and the relationship between the staggered
perturbation theory picture and the actual numerical results were further clarified in \cite{Aubin:2015dgk}.

\begin{figure}
\begin{center}
\includegraphics[width=12cm]{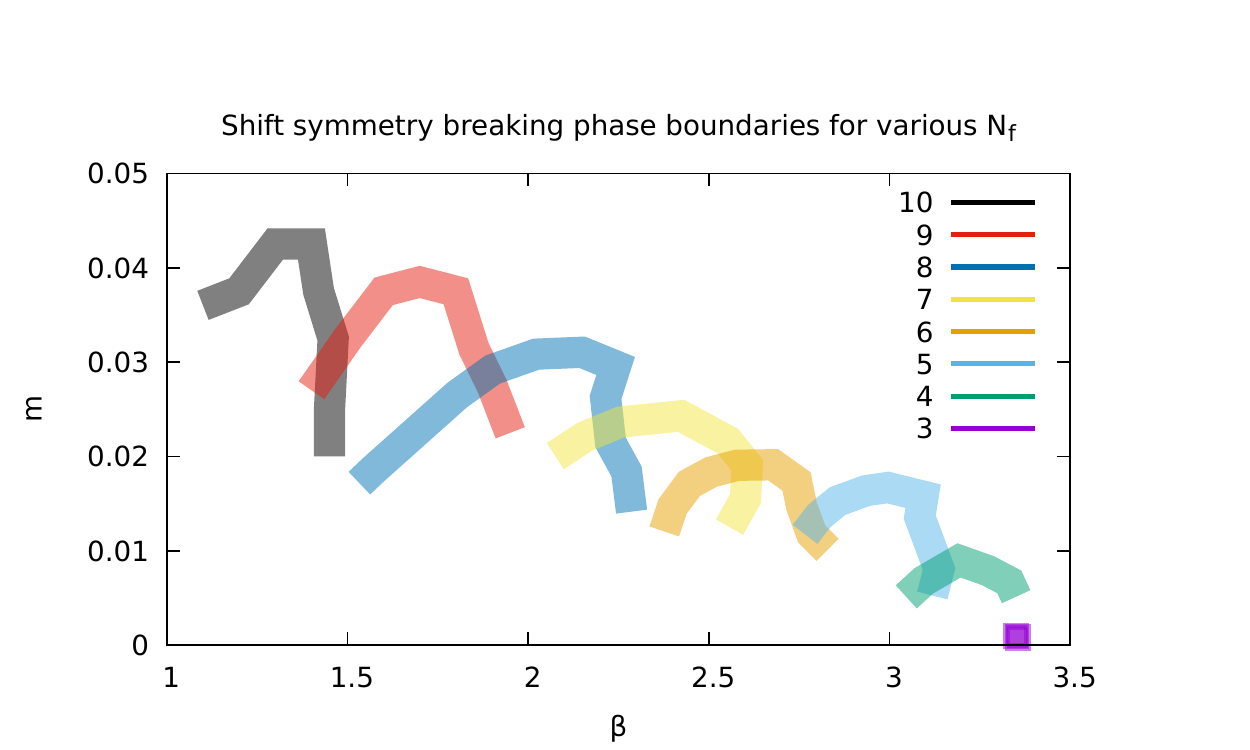} 
\end{center}
    \caption{The shift symmetry breaking phase boundaries in the $(\beta,m)$ plane. The broken phase is
    located under the curves, i.e. for small quark masses. The curves are given with different colors for
    different number of flavors.}
\label{bound}
\end{figure}

In this section we study the unphysical, shift symmetry broken phase with our particular
discretization and the full range of flavor numbers $2 \leq N_f \leq 10$ contained in both
\cite{Nogradi:2019iek} and the present work. The main conclusion will be that even though unphysical
phases do exist for $N_f > 2$ and we do map them out, our simulation points are all in the physical
phase, justifying our chiral-continuum extrapolations.

The single site shift symmetry in question is \cite{Golterman:1984cy}
\bea
\chi(x) \to \xi_\mu(x) \chi(x+{\hat\mu})\;, \qquad 
{\bar\chi}(x) \to {\bar\chi}(x+{\hat\mu}) \xi_\mu(x)\;, \qquad U_\mu(x) \to U_\mu(x+{\hat\mu})\;,
\label{tr}
\eea
where $\chi(x)$ is the staggered field at integer site $x$, ${\hat\mu}$ is the unit vector on the lattice in direction
$\mu$ and $\xi_\mu(x) = (-1)^{\sum_{\nu > \mu} x_\nu}$. In this convention the staggered signs in the
Dirac operator are $\eta_\mu(x) = (-1)^{\sum_{\nu < \mu} x_\nu}$. The staggered action is clearly
invariant under this set of transformations.

As discussed in \cite{Cheng:2011ic} a suitable order parameter for the study of the
potential spontaneous breaking of (\ref{tr}) is the difference of plaquettes on neighboring sites. 
More precisely, in terms of the plaquette $P(x)$,
\bea
\Delta_\mu P &=& \sum_{x_\mu\; even} \langle P(x+{\hat\mu}) - P(x)  \rangle\;,
\label{defp}
\eea
where the sum over the lattice involves only even $x_\mu$ coordinates. Clearly, if the sum would be over
the entire lattice $\Delta_\mu P$ would always be zero. In this way $\Delta_\mu P$ measures if
translational invariance in direction $\mu$ holds for the plaquette or not. In the physical phase, 
where translational invariance for gluonic observables is present, $\Delta_\mu P = 0$ for all $\mu$.
The unphysical phase will be signaled by $\Delta_\mu P \neq 0$ for at least one direction $\mu$.

It is a straightforward exercise to map the observable $\Delta_\mu P$ as a function of $(\beta,m)$ for the
various flavor numbers. A useful quantity to monitor is the square $\Delta_\mu P \Delta_\mu P$ involving
a sum over $\mu$. Two typical results are shown for $\Delta_\mu P \Delta_\mu P$ at fixed $\beta$ as a
function of $m$ and at fixed $m$ as a function of $\beta$ in figure \ref{deltap} with $N_f = 7$ on
$18^4$ lattices. It is
not our goal to obtain very precise values for $(\beta_c,m_c)$ corresponding to the spontaneous breaking
of translational invariance, for our purposes an estimate will suffice which can be read off from results of
the type shown in figure \ref{deltap}. A detailed finite size scaling study would be required for anything more precise.
As we will see our simulation points are so far away from the $(\beta_c,m_c)$
phase boundaries that a rough estimate is indeed sufficient.

Performing the scans on $12^4$ and $18^4$ lattices shows that volume dependence is negligible on
our level of precision. The summary of our results for the phase boundaries are shown in figure
\ref{bound} for all flavor numbers where the thickness of the boundaries include the uncertainty related
to our crude reading off of $(\beta_c,m_c)$ on fixed $18^4$ lattice volumes.

For each $N_f > 2$ a triangle shaped region corresponds to the spontaneously broken shift symmetry phase at finite
$(\beta,m)$. This triangle presumably extends down to $m = 0$ at two particular $\beta$ values. The bare
mass, above which translational symmetry is unbroken for all $\beta$ is a growing function of $N_f$ as
can be seen in figure \ref{bound}. Not surprisingly, the particular $\beta$ above which translational symmetry
is unbroken for all masses is a decreasing function of $N_f$.
At $N_f = 3$ we could not resolve the triangle shape because the
broken phase only occures for very small masses, but nevertheless could find a transition. Interestingly,
we could not detect any translational symmetry broken phase for $N_f = 2$, perhaps because no such phase
exists or perhaps because it occurs at extremely small masses.

The $(\beta,m)$ values for $N_f = 2,3,4,5,6$ which were used in the chiral-continuum extrapolations in
\cite{Nogradi:2019iek} were listed in tables 3 and 4 of said work while the same parameters are listed in
table \ref{data} for the present work with $N_f = 7,8,9,10$. Clearly, all parameters used
for the chiral-continuum extrapolations are in the physical phase and far from the $(\beta_c,m_c)$ phase
boundaries.

\section{Finite volume effects}
\label{finitevolumeeffects}

Just as in \cite{Nogradi:2019iek}, a prerequisite step before chiral-continuum extrapolations are
performed is the study of finite volume effects. The volume, measured in $m_\pi$ units, needs to be large
enough in order to suppress finite volume distortions of the ratio $m_\varrho / f_\pi$
especially because as the volume is increasing $m_\varrho$ and $f_\pi$ are moving in the opposite
direction. The
finite volume effects are thus enhancing each other in the ratio and too small volumes will lead to an
overestimation of $m_\varrho / f_\pi$. 

\begin{table}
\footnotesize
\begin{center}
\begin{tabular}{|c|c|c|c|c|c|c|c|c|c|}
\hline
$N_f$ & $\beta$ & $m$ & $L/a$ & $am_\pi$ & $af_\pi$ \\
\hline
\hline
7 & 3.00 & 0.0100 & 20 & 0.210(3) & 0.0385(6) \\
\hline
   &      &        & 24 & 0.197(1) & 0.0427(4) \\
\hline
   &      &        & 28 & 0.1913(5) & 0.0434(3) \\
\hline
   &      &        & 32 & 0.1914(5) & 0.0443(2) \\
\hline
   &      &        & $\infty$ & 0.1900(7) $\;\;$ 2.52 & 0.0444(2) $\;\;$ 1.24 \\
\hline
\hline
8 & 2.68 & 0.0103 & 20 & 0.239(3) & 0.0339(5) \\
\hline
   &      &        & 24 & 0.209(1) & 0.0396(4) \\
\hline
   &      &        & 28 & 0.1999(9) & 0.0415(3) \\
\hline
   &      &        & 32 & 0.1983(5) & 0.0417(1) \\
\hline
   &      &        & $\infty$ & 0.1964(8) $\;\;$ 0.87 & 0.0421(2) $\;\;$ 0.36 \\
\hline
\hline
9 & 2.49 & 0.0100 & 28 & 0.196(1) & 0.0294(2) \\
\hline
   &      &        & 32 & 0.181(1) & 0.0320(2) \\
\hline
   &      &        & 36 & 0.1771(9) & 0.0329(2) \\
\hline
   &      &        & 40 & 0.1756(5) & 0.0324(2) \\
\hline
   &      &        & $\infty$ & 0.1740(6) $\;\;$ 0.59 & 0.0330(1) $\;\;$ 6.04 \\
\hline
\hline
10 & 2.30 & 0.0112 & 28 & 0.228(1) & 0.0238(2) \\
\hline
   &      &        & 32 & 0.194(2) & 0.0257(3) \\
\hline
   &      &        & 36 & 0.180(1) & 0.0273(2) \\
\hline
   &      &        & 40 & 0.174(1) & 0.0277(1) \\
\hline
   &      &        & 48 & 0.1704(5) & 0.02813(8) \\
\hline
   &      &        & $\infty$ & 0.1699(6) $\;\;$ 0.45 & 0.02807(9) $\;\;$ 2.88 \\
\hline
\hline
\end{tabular}
\end{center}
\caption{Volume dependence of $m_\pi$ and $f_\pi$ and fixed lattice spacing and fermion mass, together
    with the infinite volume extrapolated results using (\ref{g1fit}) and (\ref{g1}). The $\chi^2/dof$ of
    the extrapolations are also shown.}
\label{datafinvol}
\end{table}

An upper bound on the size of finite volume effects 
sets a lower bound on $m_\pi L$ for each $N_f$.
The results in \cite{Nogradi:2019iek} have shown that this lower bound is heavily $N_f$-dependent in the
range $2 \leq N_f \leq 6$. In the present work these finite volume investigations are extended to $7 \leq
N_f \leq 10$.

The main low energy quantities $m_\pi$ and $f_\pi$ are measured at fixed lattice spacing and mass
$m$ for various lattice volumes, since these observables are expected to be the most sensitive to the
finite volume. The $m_\pi L$ dependence of these quantities are given by 
\bea
m_\pi(L) &=& m_{\pi\infty} + C_m\; g(m_{\pi\infty}L) \nn \\
f_\pi(L) &=& f_{\pi\infty} - C_f \; g(m_{\pi\infty}L)\;,
\label{g1fit}
\eea
with some $m_{\pi\infty},\; f_{\pi\infty},\; C_m$ and $C_f$ parameters. The details follow the
procedure explained in \cite{Nogradi:2019iek}, in particular we have
\bea
g(x) &=& \frac{4}{x} \sum_{n\neq 0} \frac{K_1(nx)}{n}
\label{g1}
\eea
in terms of the Bessel function $K_1$. The sum is over integers $(n_1,n_2,n_3,n_4)$ with 
$n^2 = n_1^2 + n_2^2 + n_3^2 + 4n_4^2 \neq 0$ where $\mu = 4$ corresponds to the time direction. 
The function $g(x)$ describing the finite volume effects
represents the lightest particle, the pion, going around the finite volume in all 3 space and the time direction any
number of times. The leading contribution for our geometry, where the lattice is largest in the time
direction comes from the pion going around each spatial direction once, corresponding to $n=(\pm 1, 0, 0,
0),\; (0, \pm 1, 0, 0),\; ( 0, 0, \pm 1, 0 )$. If only these terms are kept we are led to the familiar
finite volume effects given by a single exponential,
\bea
g(x) &=& 24 \sqrt{\frac{\pi}{2}} \frac{e^{-x}}{x^{3/2}} \left( 1 +
O\left(\frac{1}{x}\right)\right) + O\left(e^{-\sqrt{2}x}\right)\;.
\label{g1e}
\eea
We have repeated the finite volume fits with the above leading order single exponential expression as well and
the results did not change within statistical uncertainties hence the data can not distinguish between
the two sets of extrapolations. Note that the finite volume extrapolations (\ref{g1fit}) with either
(\ref{g1}) or (\ref{g1e}) do not depend on chiral perturbation theory at all, they hold for any massive
QFT with
$m_\pi$ taking the place of the lightest mass. In particular even if $N_f$ is inside the conformal window
but a finite mass is introduced leading to finite masses for physical excitations, finite volume effects
are still described by (\ref{g1fit}) and (\ref{g1}) or approximately (\ref{g1e}).

\begin{figure}
\begin{center}
\ig[width=7.5cm]{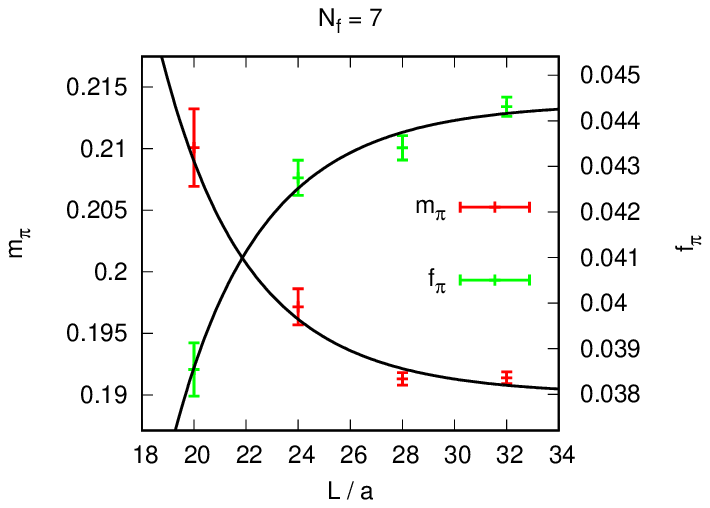} \ig[width=7.5cm]{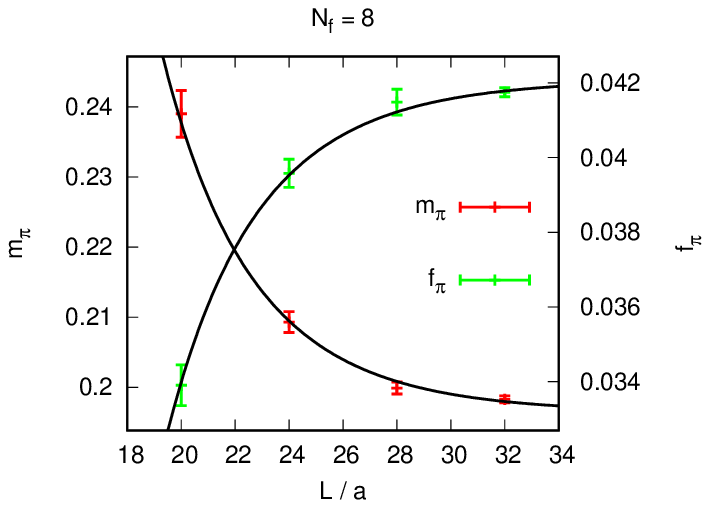} \\ 
\vspace{0.4cm} 
\ig[width=7.5cm]{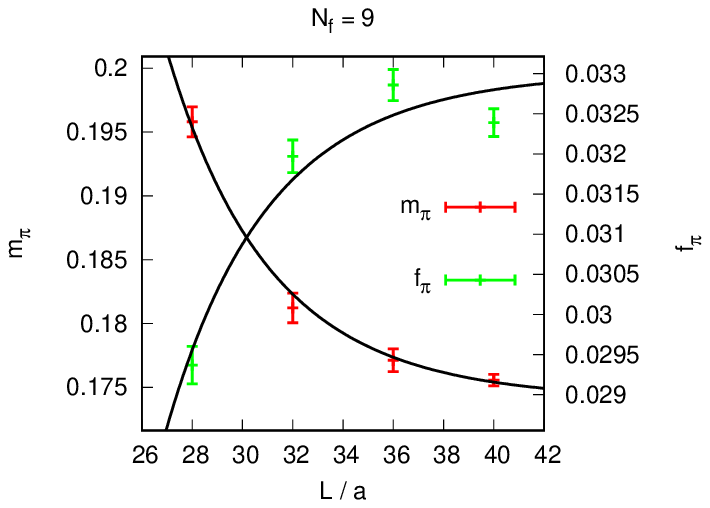} \ig[width=7.5cm]{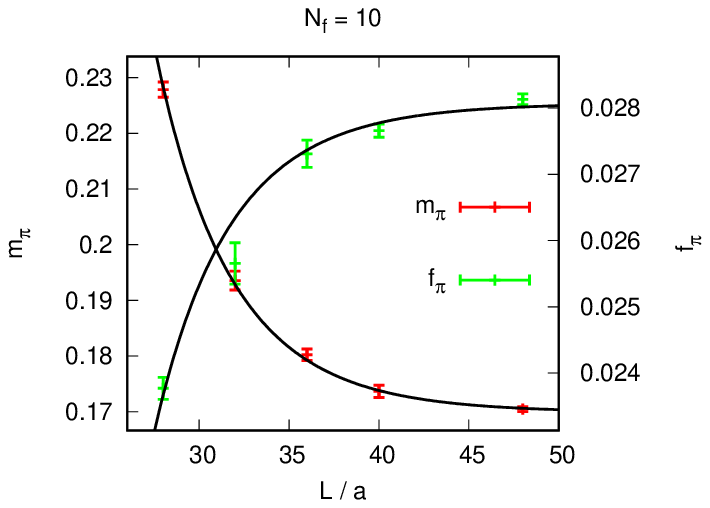}
\end{center}
\caption{Finite volume effects for $m_\pi$ and $f_\pi$ for all flavors $7 \leq N_f \leq 10$. 
Extrapolations are via (\ref{g1}) and the data is
tabulated in table \ref{datafinvol}.}
\label{finvol}
\end{figure}

The results of our fits of the type (\ref{g1fit}) with (\ref{g1}) are shown in figure \ref{finvol}.

The main conclusion from the finite volume volume study is that as $N_f$ is increasing the minimal
$m_\pi L$ required for at most $1\%$ finite volume effects needs to grow. On the full range $2 \leq N_f
\leq 10$ including the results from \cite{Nogradi:2019iek} the bounds can be interpolated by the simple
expression,
\bea
m_\pi L > 3.46 + 0.12 N_f + 0.03 N_f^2\;.
\label{mpil}
\eea
For instance at $N_f = 10$ we have $m_\pi L > 7.66$, about twice as large as the corresponding bound
at $N_f = 2$. 

Apart from the exponential finite volume effects discussed above for $f_\pi$ and $m_\pi$, there might be
further finite volume effects influencing $m_\varrho$ because of its possible decay to 2 pions. For our
simulation points $\varrho$ is however stable. 

The conclusion from this section is that our simulation results suffer from at most 1\% finite
volume effects for $N_f = 7, 8, 9$ and at most 1.5\% for $N_f = 10$, resulting in at most 3\% distortion 
in the ratio $m_\varrho / f_\pi$, well below our statistical uncertainties.


\section{Chiral-continuum extrapolation}
\label{chiralcontinuum}

\begin{figure}
\begin{center}
\includegraphics[width=6.25cm]{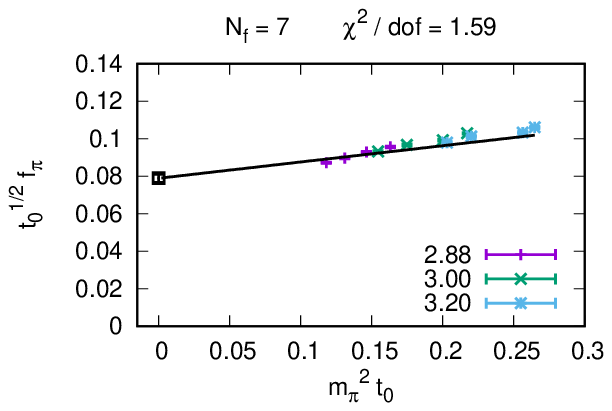} \includegraphics[width=6.25cm]{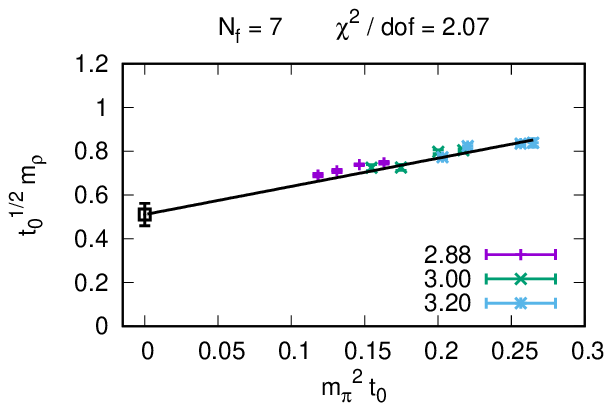} \\
\includegraphics[width=6.25cm]{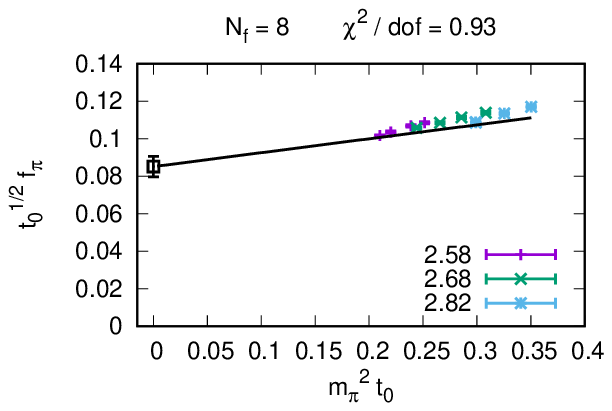} \includegraphics[width=6.25cm]{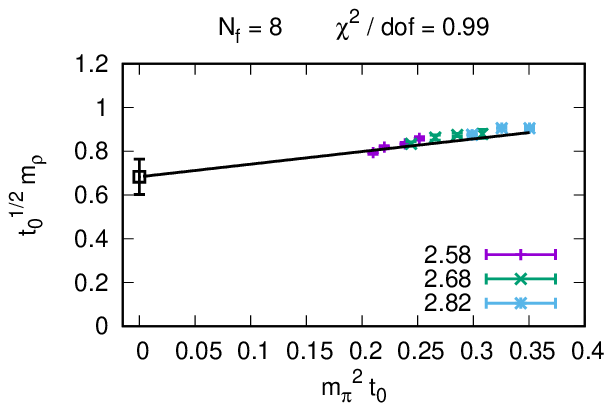} \\
\includegraphics[width=6.25cm]{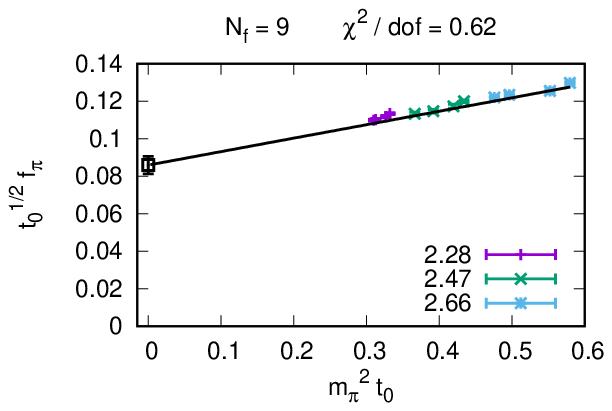} \includegraphics[width=6.25cm]{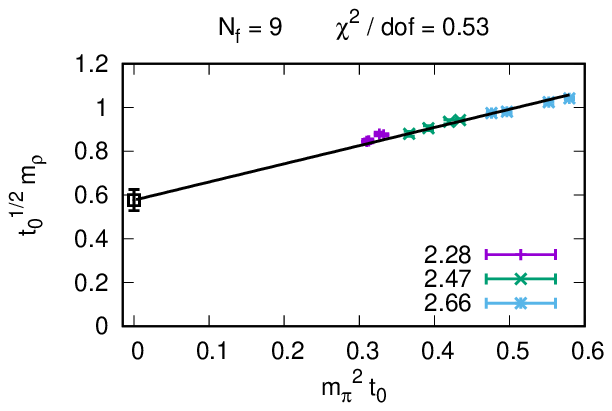} \\
\includegraphics[width=6.25cm]{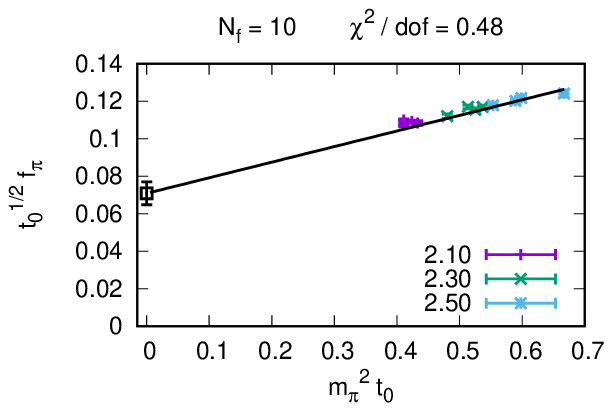} \includegraphics[width=6.25cm]{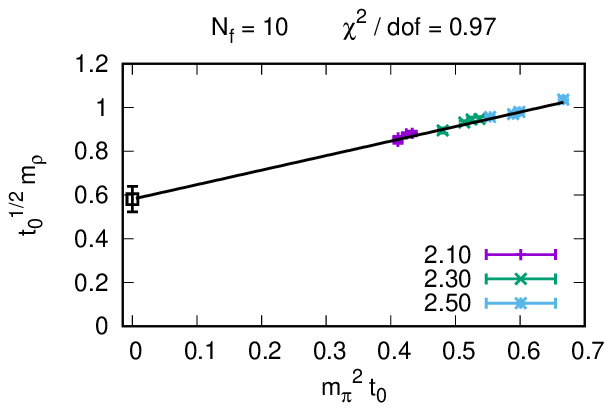} 
\end{center}
\caption{Chiral-continuum extrapolation of $f_\pi$ and $m_\varrho$ in $t_0$ units using the 4-parameter
global fit (\ref{chiralcontformula}). The $\chi^2/dof$ of the extrapolation is also shown. The solid black line
    corresponds to the resulting continuum mass dependence $C_0 + C_1 m_\pi^2 t_0,\;$ i.e.
    dropping $C_2$ and $C_3$ which are responsible for the cut-off effects. The deviations from the data
    at given bare coupling $\beta$ shown by different colors, 
    and the straight line are indicative of said cut-off effects. The absolute scale on the axis can
    not be directly compared between different flavor numbers because the definition of $t_0$ was
    $N_f$-dependent, see (\ref{t0setting}).
    }
\label{chiralcont}
\end{figure}

Apart from the observables $m_\pi, f_\pi$ and $m_\varrho$ the gradient flow scale $t_0$ was also measured
to set the scale in the chiral-continuum extrapolations. The right hand side in the definition of $t_0$
\cite{Luscher:2010iy},
\bea
\langle t_0^2 E(t_0) \rangle = c
\label{t0setting}
\eea
is in principle arbitrary, in QCD usually $c = 0.3$ is used. It is possible to choose different $c =
c(N_f)$ for
different $N_f$ though. A combination of cut-off effects, statistical uncertainty and computational
resources led to our following choices $c(7) = 0.45,\; c(8)=0.45,\; c(9)=0.40,\; c(10)=0.32$.

Once $t_0$ is measured along with our low energy quantities of interest the chiral-continuum
extrapolation is performed via
\bea
X \sqrt{t_0} = C_0 + C_1 m_\pi^2 t_0 + C_2 \frac{a^2}{t_0} + C_3 \frac{a^2}{t_0} m_\pi^2 t_0\;,
\label{chiralcontformula}
\eea
where $X = f_\pi$ or $m_\varrho$. The continuum mass dependence is given by $C_0 + C_1 m_\pi^2 t_0$ and
the two terms $C_2$ and $C_3$ parametrize cut-off effects in both the chiral limit value $C_0$ and the
slope $C_1$.

The measured data for $m_\pi, f_\pi, m_\varrho, t_0$ are shown in figure \ref{data}. The lattice geometry
was always $L^3 \times 2L$, the collected number of thermalized configurations $O(1000)$ and
every $10^{th}$ was used for measurements. For each flavor number, simulations are performed at 3 lattice
spacings with 4 masses at each. Hence the chiral-continuum extrapolations (\ref{chiralcontformula})
correspond to $dof = 8$ in each case.

\begin{figure}
\begin{center}
\includegraphics[width=8cm]{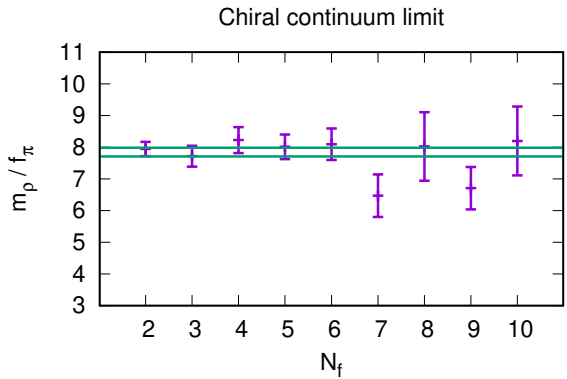}
\end{center}
\caption{The $N_f$-dependence of $m_\varrho / f_\pi$ in the chiral-continuum limit. The results with $2
    \leq N_f \leq 6$ are from \cite{Nogradi:2019iek} and $7 \leq N_f \leq 10$ corresponds to this work.
    The result of a constant fit as a function of $N_f$ is also shown.
    }
\label{nf}
\end{figure}

The chiral-continuum extrapolations are shown in figure \ref{chiralcont} and the results are tabulated in
table \ref{chiralconttable}. The full $N_f$-dependence of $m_\varrho / f_\pi$ in the chiral-continuum
limit for $2 \leq N_f \leq 10$ using also the results from \cite{Nogradi:2019iek} is shown in figure
\ref{nf}.

\begin{table}
\footnotesize
\begin{center}
\begin{tabular}{|c|c|c|c|}
\hline
$N_f$ & $f_\pi \sqrt{t_0}$ & $m_\varrho \sqrt{t_0}$ & $m_\varrho / f_\pi$ \\
\hline
\hline
    7 &    0.079(2) & 0.51(5) & 6.5(7) \\
\hline            
    8 &    0.085(5) & 0.68(8) & 8.0(1.1) \\
\hline            
    9 &    0.086(5) & 0.58(5) & 6.7(7) \\
\hline            
    10 &   0.071(6) & 0.58(6) & 8.2(1.1) \\
\hline
\end{tabular}
\end{center}
\caption{
\label{chiralconttable}
Continuum results for each $N_f$ in the chiral limit.}
\end{table}

It was observed in \cite{Nogradi:2019iek} that there is no statistically significant $N_f$-dependence in
the ratio for $2\leq N_f \leq 6$, at least on the level of precision available there. A statistically
good constant fit gave
$m_\varrho / f_\pi = 7.95(15)$. We can now repeat the constant fit on the
new range $7 \leq N_f \leq 10$ and the result is $m_\varrho / f_\pi = 7.01(40)$ with $\chi^2/dof = 0.97$,
which represents a slight $2\sigma$-decrease.
Nevertheless combining all results on the full range $2 \leq N_f \leq 10$ 
we obtain $m_\varrho / f_\pi = 7.85(14)$ with $\chi^2/dof = 1.10$ which is our final result. 
\footnote{As a consistency check we have also fitted the $N_f$-dependence as 
$m_\varrho / f_\pi = A + N_f B$ which resulted in $A = 8.17(29)$, $B = -0.083(66)$ with $\chi^2/dof =
1.04$. The fit parameter $B$ is consistent with zero on the $1.3\sigma$ level.}
Apparently,
the free value $m_\varrho / f_\pi = 1/\sqrt{3}$ at $N_f = 33/2$ is still about an order of magnitude away.

\begin{table}
\footnotesize
\begin{center}
\begin{tabular}{|c|c|c|c|c|c|c|c|c|c|}
\hline
$N_f$ & $\beta$ & $m$ & $L/a$ & $am_\pi$ & $af_\pi$ & $a m_\varrho$ & $t_0/a^2$ & $m_\pi L$ & $f_\pi L$ \\
\hline
\hline
\hline
 7 & 2.88 & 0.0149 & 24 & 0.2583(5) & 0.0612(2) & 0.478(3) & 2.45(2) & 6.20(1) & 1.469(5) \\
\hline
   &      & 0.0122 & 28 & 0.2327(4) & 0.0565(1) & 0.449(2) & 2.70(2) & 6.52(1) & 1.583(4) \\
\hline
   &      & 0.0100 & 28 & 0.2092(4) & 0.0519(1) & 0.410(4) & 2.99(2) & 5.86(1) & 1.452(4) \\
\hline
   &      & 0.0086 & 32 & 0.1932(4) & 0.04904(9) & 0.389(4) & 3.17(2) & 6.18(1) & 1.569(3) \\
\hline
\hline
   & 3.00 & 0.0147 & 28 & 0.2353(8) & 0.0520(2) & 0.406(3) & 3.92(2) & 6.59(2) & 1.455(4) \\
\hline
   &      & 0.0125 & 28 & 0.2164(4) & 0.0480(2) & 0.386(4) & 4.28(4) & 6.06(1) & 1.345(4) \\
\hline
   &      & 0.0100 & 32 & 0.1914(5) & 0.0443(2) & 0.333(6) & 4.77(3) & 6.12(2) & 1.418(7) \\
\hline
   &      & 0.0084 & 36 & 0.1738(5) & 0.0412(1) & 0.320(6) & 5.12(3) & 6.26(2) & 1.484(5) \\
\hline
\hline
   & 3.20 & 0.0115 & 36 & 0.1741(6) & 0.0359(2) & 0.284(6) & 8.74(6) & 6.27(2) & 1.292(6) \\
\hline
   &      & 0.0100 & 36 & 0.1630(8) & 0.0333(2) & 0.269(4) & 9.65(6) & 5.87(3) & 1.198(7) \\
\hline
   &      & 0.0085 & 40 & 0.1479(3) & 0.0319(2) & 0.260(4) & 10.07(9) & 5.92(1) & 1.277(7) \\
\hline
   &      & 0.0077 & 40 & 0.1403(5) & 0.0305(2) & 0.241(3) & 10.3(1) & 5.61(2) & 1.219(7) \\
\hline
\hline
\hline
 8 & 2.58 & 0.0149 & 24 & 0.2619(5) & 0.0567(2) & 0.448(4) & 3.67(3) & 6.28(1) & 1.361(5) \\
\hline
   &      & 0.0124 & 28 & 0.2355(5) & 0.0515(2) & 0.402(2) & 4.30(5) & 6.59(1) & 1.442(4) \\
\hline
   &      & 0.0099 & 28 & 0.2096(8) & 0.0463(1) & 0.366(3) & 5.01(5) & 5.87(2) & 1.296(4) \\
\hline
   &      & 0.0087 & 32 & 0.1945(5) & 0.0432(1) & 0.337(3) & 5.55(3) & 6.23(2) & 1.383(4) \\
\hline
\hline
   & 2.68 & 0.0145 & 28 & 0.2400(7) & 0.0492(2) & 0.380(8) & 5.35(5) & 6.72(2) & 1.378(5) \\
\hline
   &      & 0.0124 & 28 & 0.2207(6) & 0.0460(2) & 0.361(2) & 5.86(6) & 6.18(2) & 1.288(4) \\
\hline
   &      & 0.0103 & 32 & 0.1983(5) & 0.0417(1) & 0.332(4) & 6.76(5) & 6.34(2) & 1.336(4) \\
\hline
   &      & 0.0083 & 36 & 0.1750(5) & 0.0375(1) & 0.296(3) & 7.97(7) & 6.30(2) & 1.350(4) \\
\hline
\hline
   & 2.82 & 0.0120 & 36 & 0.1959(6) & 0.0387(2) & 0.300(3) & 9.13(8) & 7.05(2) & 1.394(6) \\
\hline
   &      & 0.0100 & 36 & 0.1770(5) & 0.0352(1) & 0.281(3) & 10.4(1) & 6.37(2) & 1.269(5) \\
\hline
   &      & 0.0080 & 36 & 0.1583(6) & 0.0314(3) & 0.253(4) & 11.9(2) & 5.70(2) & 1.132(9) \\
\hline
   &      & 0.0075 & 40 & 0.1515(5) & 0.0302(1) & 0.243(2) & 13.0(1) & 6.06(2) & 1.207(5) \\
\hline
\hline
\hline
 9 & 2.28 & 0.0164 & 28 & 0.2672(5) & 0.0526(2) & 0.406(2) & 4.66(4) & 7.48(1) & 1.472(4) \\
\hline
   &      & 0.0128 & 32 & 0.2311(5) & 0.0452(1) & 0.355(3) & 6.12(6) & 7.39(2) & 1.447(4) \\
\hline
   &      & 0.0100 & 36 & 0.1994(3) & 0.0393(1) & 0.302(3) & 7.86(9) & 7.18(1) & 1.416(4) \\
\hline
   &      & 0.0090 & 40 & 0.1875(2) & 0.03703(9) & 0.285(3) & 8.81(7) & 7.500(9) & 1.481(4) \\
\hline
\hline
   & 2.47 & 0.0140 & 32 & 0.2198(5) & 0.0400(1) & 0.315(2) & 8.98(8) & 7.03(2) & 1.281(4) \\
\hline
   &      & 0.0110 & 36 & 0.1906(5) & 0.0345(2) & 0.275(2) & 11.5(2) & 6.86(2) & 1.243(6) \\
\hline
   &      & 0.0090 & 40 & 0.1684(6) & 0.0309(2) & 0.244(2) & 13.8(2) & 6.74(2) & 1.234(7) \\
\hline
   &      & 0.0070 & 48 & 0.1442(3) & 0.0270(1) & 0.210(2) & 17.6(2) & 6.92(2) & 1.296(5) \\
\hline
\hline
   & 2.66 & 0.0200 & 28 & 0.248(1) & 0.0423(2) & 0.340(1) & 9.4(1) & 6.95(3) & 1.185(7) \\
\hline
   &      & 0.0150 & 32 & 0.2078(7) & 0.0351(2) & 0.287(2) & 12.8(2) & 6.65(2) & 1.124(7) \\
\hline
   &      & 0.0120 & 40 & 0.1785(8) & 0.0313(2) & 0.249(2) & 15.6(3) & 7.14(3) & 1.250(7) \\
\hline
   &      & 0.0098 & 48 & 0.1568(5) & 0.0277(2) & 0.222(2) & 19.3(2) & 7.53(2) & 1.331(7) \\
\hline
\hline
\hline
10 & 2.10 & 0.0165 & 32 & 0.2423(7) & 0.0416(2) & 0.326(2) & 7.00(7) & 7.75(2) & 1.330(5) \\
\hline
   &      & 0.0126 & 36 & 0.2037(7) & 0.0342(1) & 0.275(1) & 10.2(1) & 7.33(2) & 1.230(5) \\
\hline
   &      & 0.0100 & 40 & 0.1801(9) & 0.0297(2) & 0.241(2) & 13.4(3) & 7.20(3) & 1.186(6) \\
\hline
   &      & 0.0081 & 48 & 0.1535(6) & 0.02567(9) & 0.202(2) & 17.4(2) & 7.37(3) & 1.232(4) \\
\hline
\hline
   & 2.30 & 0.0185 & 32 & 0.2386(7) & 0.0390(1) & 0.310(1) & 9.0(1) & 7.63(2) & 1.247(5) \\
\hline
   &      & 0.0142 & 36 & 0.2005(9) & 0.0320(1) & 0.259(2) & 13.4(2) & 7.22(3) & 1.152(5) \\
\hline
   &      & 0.0112 & 40 & 0.174(1) & 0.0277(1) & 0.227(1) & 17.4(3) & 6.95(4) & 1.106(4) \\
\hline
   &      & 0.0091 & 48 & 0.1502(9) & 0.0243(2) & 0.194(2) & 21.3(4) & 7.21(4) & 1.166(7) \\
\hline
\hline
   & 2.50 & 0.0233 & 28 & 0.2584(9) & 0.0393(3) & 0.328(2) & 10.0(2) & 7.23(3) & 1.100(8) \\
\hline
   &      & 0.0178 & 36 & 0.2106(5) & 0.0332(2) & 0.267(2) & 13.5(3) & 7.58(2) & 1.194(8) \\
\hline
   &      & 0.0141 & 40 & 0.1800(7) & 0.0282(1) & 0.227(2) & 18.2(3) & 7.20(3) & 1.128(6) \\
\hline
   &      & 0.0114 & 48 & 0.1571(6) & 0.0249(1) & 0.202(1) & 22.4(2) & 7.54(3) & 1.195(7) \\
\hline
\hline
\end{tabular}
\end{center}
\caption{Data used for the chiral-continuum extrapolations. The temporal extent of the lattices were always twice $L/a$.}
\label{data}
\end{table}

\section{Conclusion}
\label{conclusion}

In this work we continued our study of the ratio $m_\varrho / f_\pi$ in the chiral-continuum limit. 
Constant fits as a function of $N_f$ on the two ranges $2 \leq N_f \leq 6$ and $7 \leq N_f \leq 10$ show
a decrease on the $2\sigma$-level but a constant fit on the full range $2\leq N_f \leq 10$ is still a
statistically acceptable result and leads to $m_\varrho / f_\pi = 7.85(14)$. 

The main conclusion is the reinforcement of the picture arising from \cite{Nogradi:2019iek}, namely that
$m_\varrho / f_\pi$ is a robust quantity once the gauge group is fixed and does not depend much, if at 
all, on the fermion content. Applied to composite Higgs models inspired by strong dynamics, this
would mean that a potential measurement of a new so far unobserved vector resonance inherent in these types
of models, would not select the flavor number. The measured vector mass would rather place constraints
on the gauge group \cite{Nogradi:2019auv}.

Our ratio has a well-defined meaning in the chiral limit both inside and outside the conformal window.
If the free value $m_\varrho / f_\pi = 1 / \sqrt{3} = 0.577$ is to be reached at $N_f = 16.5$, an order
of magnitude drop ought to take place beyond $N_f = 10$. If the trends of finite volume effects follow
(\ref{mpil}) in any sense, the $N_f > 10$ simulations will be very challenging. It would be most
interesting to work out the perturbative corrections to $1/\sqrt{3}$ close to the upper end of the conformal
window, i.e. not much below $N_f = 16.5$ where perturbation theory is reliable. Hopefully the full range
$2\leq N_f \leq 16$ can then be covered by a combination of non-perturbative simulations and perturbative
results. The onset of the conformal window would probably leave some sort of imprint on the flavor
dependence of the ratio, a subject we leave for future work.

\section*{Acknowledgements}

DN would like to thank very useful discussions with Stephan Durr and Sandor Katz.
The simulations were carried out on the GPU clusters of Eotvos University Budapest and University of
Wuppertal and at HLRS in Stuttgart, Germany.
This work was in part supported by the Hungarian National Research, 
Development and Innovation Office (NKFIH) grant KKP126769.

\end{document}